\documentclass[sigconf]{acmart}
\usepackage{multirow}
\usepackage{subfigure}
\usepackage{utfsym}
\usepackage{color}
\usepackage{balance}

\AtBeginDocument{%
  \providecommand\BibTeX{{%
    \normalfont B\kern-0.5em{\scshape i\kern-0.25em b}\kern-0.8em\TeX}}}

\copyrightyear{2023}
\acmYear{2023}
\setcopyright{acmlicensed}
\acmConference[MM '23] {Proceedings of the 31st ACM International Conference on Multimedia}{October 29--November 3, 2023}{Ottawa, ON, Canada.}
\acmBooktitle{Proceedings of the 31st ACM International Conference on Multimedia (MM '23), October 29--November 3, 2023, Ottawa, ON, Canada}
\acmPrice{15.00}
\acmISBN{979-8-4007-0108-5/23/10}
\acmDOI{10.1145/3581783.3612252}

\acmSubmissionID{2403}

\begin{document}

\title{Knowledge Prompt-tuning for Sequential Recommendation}

\author{Jianyang Zhai}
\affiliation{%
  \institution{Sun Yat-sen University}
  \city{Guangzhou}
  \country{China}
}
\affiliation{%
  \institution{PengCheng Laboratory}
  \city{Shenzhen}
  \country{China}
}
\email{zhaijy01@pcl.ac.cn}

\author{Xiawu Zheng}
\authornote{Corresponding authors.}
\affiliation{%
  \institution{PengCheng Laboratory}
  \city{Shenzhen}
  \country{China}
}
\email{zhengxw01@pcl.ac.cn}

\author{Chang-Dong Wang}
\authornotemark[1]
\affiliation{%
  \institution{Sun Yat-sen University}
  \institution{Guangdong Provincial Key Laboratory of Intellectual Property and Big Data}
  \city{Guangzhou}
  \country{China}
}
\email{changdongwang@hotmail.com}


\author{Hui Li}
\affiliation{%
 \institution{School of Informatics, Xiamen University}
 \city{Xiamen}
 \country{China}}
\email{hui@xmu.edu.cn}

\author{Yonghong Tian}
\affiliation{%
  \institution{Peking University}
  \city{Beijing}
  \country{China}
}
\affiliation{%
  \institution{PengCheng Laboratory}
  \city{Shenzhen}
  \country{China}
}
\email{yhtian@pku.edu.cn}

\renewcommand{\shortauthors}{Jianyang Zhai, Xiawu Zheng, Chang-Dong Wang, Hui Li, \& Yonghong Tian}

\begin{abstract}
Pre-trained language models (PLMs) have demonstrated strong performance in sequential recommendation (SR), which are utilized to extract general knowledge. However, existing methods still lack domain knowledge and struggle to capture users' fine-grained preferences. Meanwhile, many traditional SR methods improve this issue by integrating side information while suffering from information loss. To summarize, we believe that a good recommendation system should utilize both general and domain knowledge simultaneously. Therefore, we introduce an external knowledge base and propose Knowledge Prompt-tuning for Sequential Recommendation (\textbf{KP4SR}). Specifically, we construct a set of relationship templates and transform a structured knowledge graph (KG) into knowledge prompts to solve the problem of the semantic gap. However, knowledge prompts disrupt the original data structure and introduce a significant amount of noise. We further construct a knowledge tree and propose a knowledge tree mask, which restores the data structure in a mask matrix form, thus mitigating the noise problem. We evaluate KP4SR on three real-world datasets, and experimental results show that our approach outperforms state-of-the-art methods on multiple evaluation metrics. Specifically, compared with PLM-based methods, our method improves NDCG@5 and HR@5 by \textcolor{red}{40.65\%} and \textcolor{red}{36.42\%} on the books dataset, \textcolor{red}{11.17\%} and \textcolor{red}{11.47\%} on the music dataset, and \textcolor{red}{22.17\%} and \textcolor{red}{19.14\%} on the movies dataset, respectively. Our code is publicly available at the link: \href{https://github.com/zhaijianyang/KP4SR}{\textcolor{blue}{https://github.com/zhaijianyang/KP4SR}.}
\end{abstract}




\maketitle

\begin{figure}[t]
  \includegraphics[width=\columnwidth]{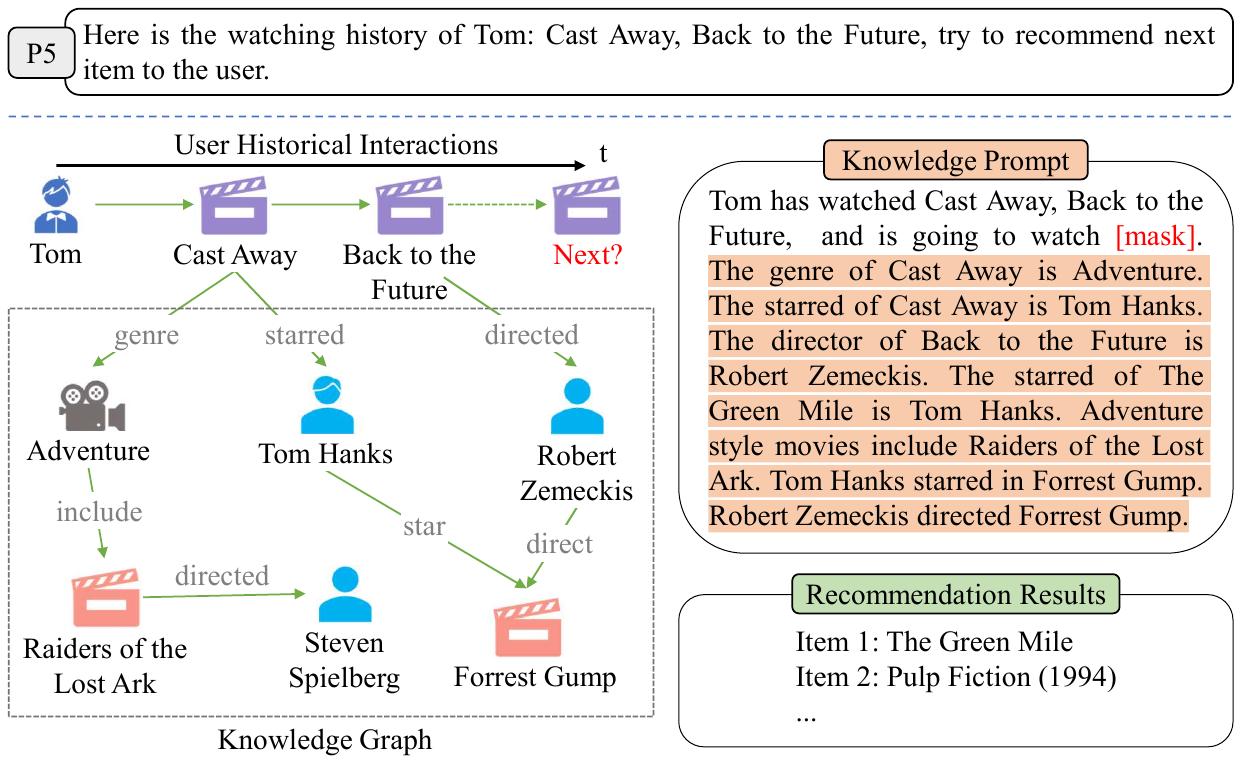}
  \caption{Comparison between P5 \cite{geng2022recommendation} (up) and our KP4SR (down), which introduces domain knowledge. We introduce an external knowledge graph and transform it into knowledge prompts to bridge the semantic gap between textual and structured data, which enriches the semantic features of items and improves the accuracy and explainability of recommendation results.}
  \label{fig:graph2text}
\end{figure}

\section{Introduction}

Recommendation systems aim to suggest desirable items among an extensive item collection to target users, thereby alleviating the problem of information overload on the internet. Traditional recommendation algorithms, such as collaborative filtering \cite{hu2008collaborative}, only mine the static correlation between users and items, ignoring the dynamic changes in user preferences implied in the historical interaction sequence. Therefore, the sequential recommendation that mines the dynamic changes in user preferences has become an important research direction in recommendation systems.

Some early studies modeled the historical behavior sequences of users using Markov chains \cite{shani2005mdp, rendle2010factorizing}, but they often struggle to handle complex sequential patterns. With the development of deep learning, deep neural networks have achieved great success in sequential recommendation \cite{hidasi2015session, tang2018personalized}. Among them, models based on self-attention mechanisms, which adaptively assign weights to user interaction sequences, have become competitive mainstream solutions \cite{kang2018self, sun2019bert4rec, de2021transformers4rec}. However, most of these methods only model the IDs of users and items, considering only the user's sequential preferences, and cannot capture the user's fine-grained preferences.

Many studies have introduced side information into SR \cite{zhang2019feature, zhou2020s3, huang2018improving} to address these issues. Some methods extract item attribute information \cite{zhang2019feature, zhou2020s3} and fuse them at different stages. For example, FDSA \cite{zhang2019feature} uses different self-attention blocks to encode items and side information and only fuses their representations in the final stage. $\operatorname{S^3Rec}$ \cite{zhou2020s3} integrates side information through pre-training. However, inefficient feature fusion methods may result in the loss of useful information. Therefore, these methods focus on finding efficient solutions for fusing item embeddings and side information embeddings. There are also some methods that introduce external KG to assist the recommendation process \cite{huang2018improving, huang2019explainable} and learn entity and relation embeddings through knowledge graph embedding (KGE) \cite{bordes2013translating, lin2015learning} techniques. However, the primary optimization objective of KGE is the completion or prediction of edges in the KG rather than recommendation tasks.
Overall, although side information can benefit recommendations, how to effectively incorporate side information into the recommendation process is still a challenging and unresolved problem.

Recently, PLMs have achieved great success in natural language processing (NLP) \cite{brown2020language, zeng2021pangu}. By learning general knowledge from large corpora through self-supervised tasks, many researchers have utilized PLMs to solve recommendation tasks \cite{liu2023pre, geng2022recommendation, wang2022towards}. For example, PEPLER \cite{li2022personalized} used GPT-2 \cite{radford2019language} to generate more natural recommendation explanations by treating users and items as personalized prompts. P5 \cite{geng2022recommendation} transformed the recommendation task into an NLP task and unified multiple recommendation tasks, such as SR, in one framework using the T5 \cite{raffel2020exploring} model. Although these methods have achieved good results, PLM-based RS methods still face difficulties in capturing complex user preferences because PLMs lack domain knowledge regarding users and items. Therefore, introducing domain knowledge is necessary when using PLMs to solve recommendation tasks. A straightforward and simple approach is to describe domain knowledge using natural language text and then use the powerful reasoning ability of PLMs to improve recommendation performance, as shown in Figure \ref{fig:graph2text}. However, there are two challenges with this approach: 1) How to convert structured knowledge graphs into text sequences. 2) Converting into text sequences may destroy the original data structure and how to deal with the noise caused by irrelevant entities and relationships.

Inspired by prompt learning in NLP \cite{brown2020language,ye2022ontology}, we propose Knowledge Prompt-tuning Sequential Recommendation (KP4SR) to overcome the above challenges. Specifically, we improve P5 \cite{geng2022recommendation} by designing a masked personalized prompt (MPP) template set to convert the SR task into a pre-training task, which not only accelerates the convergence speed but also improves the model performance. Then, we design a set of relation templates to convert triplets into triplet prompts, which are combined to form knowledge prompts (KP). Finally, we propose prompt denoising (PD), constructing a knowledge tree and a knowledge tree mask to eliminate the mutual influence of irrelevant triplets. Extensive experimental results show that our method outperforms state-of-the-art methods on multiple metrics in three real-world datasets.

\begin{itemize}
    \item We propose KP4SR, which, to the best of our knowledge, is the first work that transforms knowledge graphs into knowledge prompts to improve SR performance.
    \item We construct KP, which addresses the problems of semantic difference between structured knowledge data contained in the KG and the sequential text data used by PLMs and allows for easy utilization of high-order information from the KG.
    \item We propose PD, which mitigates knowledge noise by restoring the KG data structure in the form of a mask matrix.
    \item We conduct extensive experiments on three datasets, and the results demonstrate the effectiveness of our method. In addition, ablation experiments show that transforming the SR task into an NLP task still follows the general pattern of NLP, which indicates the great research prospects and research value of PLMs in improving the performance of recommendation systems.
\end{itemize}

\begin{figure*}[t]
  \centering
  \includegraphics[width=0.96\textwidth]{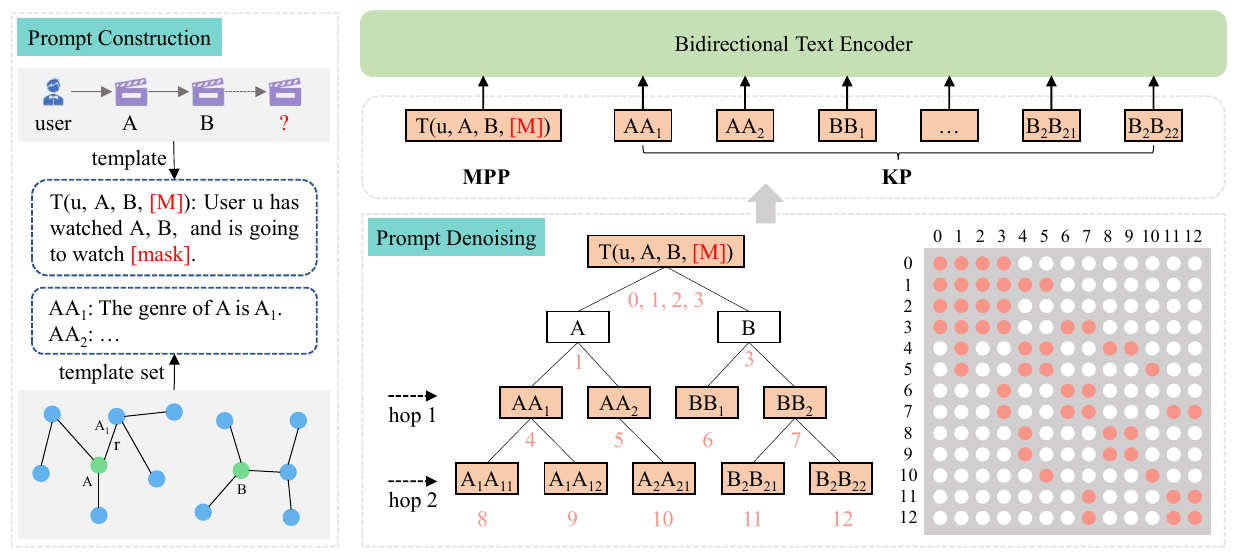}
  \caption{The overall framework of KP4SR. Firstly, we use a masked personalized prompt (MPP) template to transform the user-item interaction sequence into MPP (top-left). Then, we use relationship templates to transform the knowledge graph into a knowledge prompt (KP) (bottom-left). Finally, we construct a knowledge tree and propose a knowledge tree mask to alleviate the problem of knowledge noise (right).}
  \label{fig:framework}
\end{figure*}

\section{Related work}

\subsection{Sequential Recommendation}
Early studies utilized Markov Chains \cite{shani2005mdp, rendle2010factorizing} to model users' historical interaction sequences, but they often struggle with complex sequential patterns. Later, deep neural networks have shown strong performance in SR. GRU4Rec \cite{hidasi2015session} leverages recurrent neural networks to encode user interaction sequences into hidden states to improve recommendation performance. Caser \cite{tang2018personalized} uses horizontal and vertical convolutional filters to learn multi-level patterns and user preferences. Attention-based approaches allocate different weights to determine the relevance between users' historical interactions and target items, capturing users' dynamic preferences, such as SASRec \cite{kang2018self}, BERT4Rec \cite{sun2019bert4rec}, and Transformers4Rec \cite{de2021transformers4rec}. CLSR \cite{zheng2022disentangling} and SLi-Rec \cite{yu2019adaptive} improve recommendation performance by modeling users' long-term and short-term preferences. CL4SRec \cite{xie2022contrastive} introduces contrastive learning, which learns better sequence representations through self-supervised signals at the user behavior sequence level. However, they only model the IDs of users and items and cannot capture users' fine-grained preferences.

\subsection{Side Information for SR}
Many studies have used side information to improve SR, which mainly includes item attribute information and external knowledge base. For example, FDSA \cite{zhang2019feature} encodes items and side information using different self-attention blocks, and integrates their representations in the final stage. $\operatorname{S^3Rec}$ \cite{zhou2020s3} designs four auxiliary self-supervised objectives, utilizing the principle of maximum mutual information (MIM) to learn the correlation between attributes, items, sub-sequences, and sequences. DIF-SR \cite{xie2022decoupled} believes that early integration limits the expressive power of the attention matrix and the flexibility of gradients. It transfers the side information from the input to the attention layer and then decouples the attention calculation between various side information and item representations. KSR \cite{huang2018improving} integrates the RNN-based network and Key-Value Memory Network (KV-MN) together, and uses KG to capture users' fine-grained preferences. DHIMN \cite{xie2021sequential} applies a message-passing layer based on Dynamic Heterogeneous Information
Networks (DHIN) to capture advanced semantic knowledge in the KG, but ignores the heterogeneous information of item relationships in the KG. Compared to them, our method converts KG into text sequences and utilizes the powerful ability of PLMs to improve SR.

\subsection{PLMs for Recommendation}

PLMs have achieved tremendous success in natural language processing (NLP) \cite{devlin2018bert, brown2020language, zeng2021pangu}, and many researchers have started to use PLMs to solve recommendation tasks \cite{liu2023pre}. PEPLER \cite{li2022personalized} employs GPT-2 \cite{radford2019language} to generate personalized explanations for recommendations by using users and items as personalized prompts. METER \cite{geng2022improving} further incorporates visual information to improve the quality of recommendation explanations. In addition, SpeedyFeed \cite{xiao2022training} and NRMS \cite{wu2021empowering} utilize PLMs to enhance news recommendations. P5 \cite{geng2022recommendation} transforms recommendation tasks into NLP tasks and unifies multiple recommendation tasks in one framework using T5 \cite{raffel2020exploring}. Some works propose knowledge-enhanced dialogue recommendation systems and use PLMs to generate smoother conversations \cite{wang2022recindial, wang2022towards}. However, they used Graph Convolutional Networks (GCN) to model KG information, which faces the semantic gap that exists in natural language.

\subsection{Prompt Learning}

Prompt learning involves designing prompts for specific tasks to reframe downstream tasks as pre-training tasks, thus addressing the gap between pre-training tasks and downstream targets \cite{gu2021ppt, liu2023pre}. Early research relied on manually crafted discrete prompts to guide pre-training language models \cite{brown2020language, raffel2020exploring}. Recently, many works have focused on automatically generating discrete prompts for specific tasks \cite{gao2020making, jiang2020can}. However, these methods still rely on generative models or complex rules to control prompt quality. In contrast, some works propose using learnable continuous prompts that can be directly optimized \cite{lester2021power, li2021prefix}. Some researchers have incorporated prompt learning into recommendation systems, designing specific personalized prompts for different tasks, such as PEPLER \cite{li2022personalized} and M6-Rec \cite{cui2022m6}. Our KP4SR is an improvement on P5\cite{geng2022recommendation}, and their personalized prompts do not convert the recommendation task into a pre-training task, which is data inefficient.

\section{Problem definition}

We first introduce the symbols used in this paper. A typical recommendation scenario usually consists of a user set $\mathcal{U}$ and an item set $\mathcal{V}$. By sorting the interactions between users and items by timestamps, we can obtain the interaction sequence $S_u$ of user $u$, which can be represented as $S_u = \{ v_1^u, v_2^u, ... , v_{|u|}^u\}$, where $|u|$ denotes the length of the sequence, $u \in \mathcal{U}$ and $v \in \mathcal{V}$. Our goal is to predict the next item $v_{|u|+1}^u$ that the user is likely to interact with. To describe our method more clearly, we omit the subscripts and superscripts and simply define the user's interaction sequence as $\{A, B, C, D, E, ...\}$.

KG is a structured knowledge base that contains a set of triples, which can be defined as $\mathcal{K} \mathcal{G}=\{(h, r, t) \mid h, t \in \mathcal{E}, r \in \mathcal{R}\}$, where $\mathcal{E}$ is the set of entities and $\mathcal{R}$ is the set of relations. A triple $(h, r, t)$ represents a connection between the head entity $h$ and the tail entity $t$ through the relation $r$. We assume that each item in the recommendation system has a corresponding entity connection, i.e., $\mathcal{V} \in \mathcal{E}$. Therefore, for each item $v$, we can obtain an $n$-hop knowledge subgraph $G_v^n$ centered on $v$.

Our goal is to incorporate domain knowledge from KG into PLMs to mine users' complex preferences. For instance, given a basic input sample: \textit{"Tom has watched Cast Away, Back to the Future, and is going to watch [mask]."}, we need to input the relevant KG information, such as \textit{(Cast Away, film.genre, Adventure)}. Therefore, the first challenge we face is how to input structured KG into PLMs while bridging the semantic gap between sequence text and structured knowledge. The second challenge is how to alleviate knowledge noise, as not all knowledge is helpful, and irrelevant and noisy knowledge can affect model performance \cite{liu2020k}. In Section \ref{Methodology}, we will elaborate on our method.


\section{Methodology}
\label{Methodology}

\subsection{Overview}

We propose KP4SR, which transforms a structured KG into knowledge prompts to improve SR. It mainly consists of two modules: prompt construction module (Section \ref{prompt_construction}) and prompt denoising module (Section \ref{prompt_denoising}). The overall framework is shown in Figure \ref{fig:framework}.

The prompt construction module consists of the masked personalized prompt (MPP) and knowledge prompt (KP). Firstly, we construct a set of MPP templates by converting the user-item interaction sequence into MPP. This allows us to transform the recommendation task into a natural language cloze task. Then, we construct a set of relationship templates to convert triples into triple prompts and further combine these triple prompts to form KP. This enables us to transform structured KG into text sequences. For the prompt denoising module, we first construct a knowledge tree based on the MPP and triple prompts, and then design the knowledge tree mask to alleviate the noise problem caused by irrelevant triple prompts. We will provide more details on the method in the following sections.


\subsection{Prompts Construction}
\label{prompt_construction}

Prompt learning has achieved great success in NLP, and many researchers have applied it to recommendation systems \cite{li2022personalized, wang2022towards}. In this section, we construct a collection of MPP templates and relationship templates, which transform recommendation data and KG into textual prompts. This not only eliminates semantic differences between them but also utilizes generic knowledge in PLMs.

\begin{figure}[t]
  \includegraphics[width=\columnwidth]{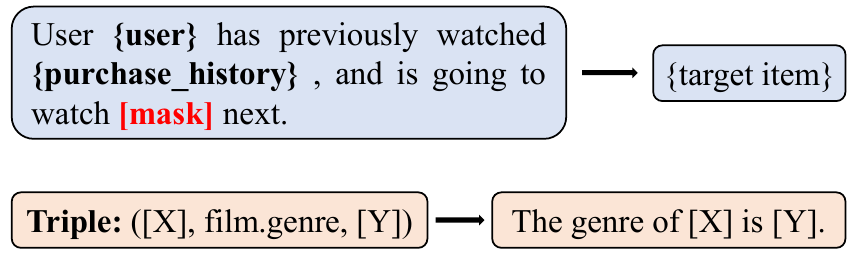}
	\caption{Examples of prompt templates. We can fill the data of user-item interaction sequences into templates to obtain MPP (up). For the relation template, we can obtain a triple prompt by replacing [X] and [Y] with the head entity and tail entity, respectively (down).}
	\label{fig:template}
\end{figure}


\subsubsection{Masked personalized prompts}

In recommendation systems, personalized prompts refer to personalized fields for different users and items \cite{geng2022recommendation, li2022personalized}. Inspired by P5 \cite{geng2022recommendation}, we construct a collection of MPP templates for SR, which helps the model discover various aspects of users and items. Specifically, MPP can transform the recommendation task into a pre-training task, namely a cloze task, as shown in Figure \ref{fig:template}. For a user $u$ and his/her interaction sequence $\{A, B, C, D, E, F\}$, we can fill in the corresponding fields in the template to obtain: \textit{User u has previously watched \{A, B, C, D, E\}, and is going to watch [mask] next.} Here, [mask] is the next item to be predicted, i.e., the target item $F$.

By constructing input-output pairs using MPP, we can leverage PLMs to extract rich semantics into user and item tokens, which will help capture users' dynamic preferences. MPP can transform the recommendation task into a pre-training task, improving task performance when downstream task data is sparse. In addition, it can also increase data utilization efficiency and accelerate model convergence.

\subsubsection{Knowledge prompts}

Using KG as side information in recommendation systems can significantly improve their performance. However, as structured data, KG cannot be directly input into PLMs, and there is a semantic gap between structured KG and sequence text. Therefore, we propose transforming the KG into a text sequence to address these issues.

For a triple $(h, r, t)$, where $h$ represents the head entity, $r$ represents the relation, and $t$ represents the tail entity, we manually design a relation template for each relation $r \in \mathcal{R}$ to express the semantics of the corresponding triple. For example, in Figure \ref{fig:template}, we design a template for the relation \textit{film.genre}: \textit{The genre of [X] is [Y].} Then for the triple \textit{(Cast Away, film.genre, Adventure)}, we replace \textit{[X]} and \textit{[Y]} with the head and tail entities, respectively, to obtain a basic triple prompt: \textit{"The genre of Cast Away is Adventure."}.


For an entity $h$, we define the set of triples $S_1=\{(h, r_1, t_1), (h, r_2, \\ t_2), ...\}$ with $h$ being the head entity as the 1-hop triple set of entity $h$, where $\{t_1, t_2, ...\}$ is the set of tail entities of entity $h$. 
Additionally, tail entities can also be treated as head entities, and they can have multiple relationships and tail entities as well. In $S_1$, there are multiple tail entities $\{t_1, t_2, ...\}$. For tail entity $t_1$, its set of 1-hop triplets can be represented as $S_{t1}=\{(t_1, r_{11}, t_{11}), (t_1, r_{12}, t_{12}), ...\}$. For tail entity $t_2$, its set of 1-hop triplets can be represented as $S_{t2}=\{(t_2, r_{21}, t_{21}), (t_2, r_{22}, t_{22}), ...\}$, and so on. Then, The set of 2-hop triplets for entity $h$ can be represented as $S_2=\{S_{t1}, S_{t2}, ...\}$.
We convert each triple into a triple prompt, which can be used to generate multi-hop triple prompts for entity $h$. We then query with the multi-hop triple prompts for each item and combine them into a text sequence to construct KP.

By constructing relation templates, we can transform structured KG into text sequences to extract multi-level information simply. 

\subsubsection{Fused Prompt}
After obtaining MPP and KP, we directly concatenate them as the input of PLM. Specifically, MPP can be represented as: $X_d=\left\{x_1, x_2, ..., [mask], ..., x_m\right\}$, where $x_i$ is the $i$-th token of the text sequence, $m$ represents the length of the text in tokens, and $[mask]$ represents the next item to be predicted. KP can be represented as: $X_k=\left\{x_1, x_2, ..., x_n\right\}$. The final input is:
\begin{equation}
    X_{prompt}=[SPE]X_d[SPE]X_k[SPE].
\end{equation}
Here, $[SPE]$ denotes a special token.

By inputting the MPP and KP into PLMs, we can integrate the recommendation task into a full language environment and use the powerful ability of PLMs to extract users' fine-grained preferences.

\subsection{Prompt Denoising}
\label{prompt_denoising}

Converting KG to knowledge prompts can disrupt the original data structure and introduce a large amount of irrelevant and noisy knowledge. For example, the two triple prompts \textit{"The genre of $h_1$ is $t_1$."} and \textit{"The director of $h_2$ is $t_2$."} have different head and tail entities and are not logically or semantically related. The mutual influence between them will generate knowledge noise. Especially when the prompts contain multi-hop triple prompts, this noise will even be amplified. To alleviate the noise problem, we use MPP and triple prompts to construct a knowledge tree, and then propose the knowledge tree mask for denoising.

\subsubsection{Konwledge tree construction}

To have a clear understanding of the structure and semantic relationships between triple prompts, we construct a knowledge tree as shown in Figure \ref{fig:framework}. 

The root node of the knowledge tree is the MPP, which contains multiple items that the user has interacted with. Therefore, the knowledge tree has multiple knowledge subtrees. Each knowledge subtree is composed of an entity and its multi-hop triple prompts. 
Suppose that the MPP contains two items corresponding to entities $A$ and $B$, then the knowledge tree consists of two knowledge subtrees, namely $subTree(A)$ and $subTree(B)$. Suppose $XX_i$ is a triple prompt consisting of $(X, r, X_i)$, then the root node of knowledge subtree $Tree(A)$ is entity $A$, and the child nodes of $A$ are its 1-hop triple prompts, $AA_1$ and $AA_2$. The 2-hop triple prompts of entity $A$ are the 1-hop triple prompts of entities $A_1$ and $A_2$, namely, $A_1A_{11}$, $A_1A_{12}$, and $A_2A_{21}$.
For example, in Figure\ref{fig:graph2text}, the movie "Cast Away" can be represented by two triplets: (Cast Away, genre, Adventure) and (Cast Away, starred, Tom Hanks). We use $A$ to represent "Cast Away", $A_1$ to represent "Adventure", and $A_2$ to represent "Tom Hanks." By introducing the relationship templates "The genre of [X] is [Y]." and "[X] starring [Y].",  we obtain two triplet prompts for $A$: "$AA_1$: The genre of Cast Away is Adventure." and "$AA_2$: Cast Away starring Tom Hanks.". $AA_1$ and $AA_2$ are 1-hop triplet prompts for $A$.

Traversing the knowledge tree in a hierarchical manner yields a sequence of prompts composed of personalized and knowledge prompts.

\subsubsection{Konwledge tree mask}

\begin{figure}[t]
  \centering
  \includegraphics[width=\columnwidth]{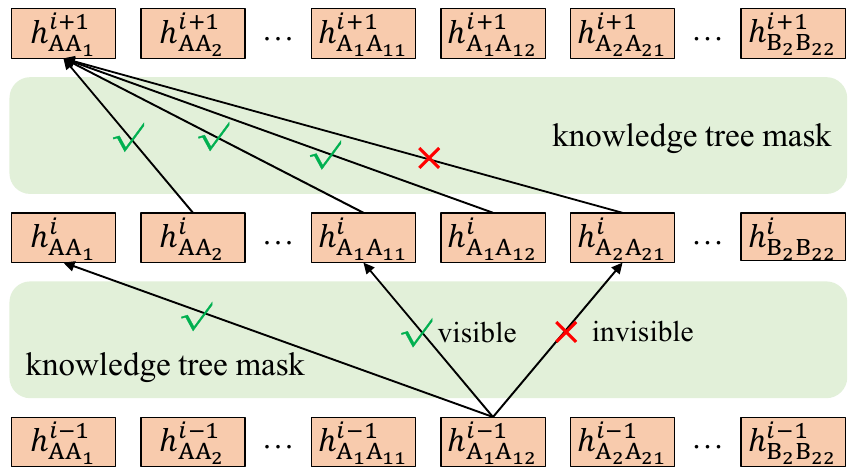}
  \caption{Illustration of prompt denoising. For any two triple prompts, if they contain the same entities, they are visible to each other; otherwise, they are invisible to each other.}
  \label{fig:visible}
\end{figure}

The knowledge tree presents the logical and semantic relationships between MPP and triple prompts. Triple prompts without logical and semantic relationships will generate much noise, and we should limit their mutual influence. For this purpose, we propose a knowledge tree mask mechanism.

Specifically, for the input sequence composed of MPP and KP, we use the mask matrix to limit the mutual influence of input tokens, as shown in Figure \ref{fig:framework}. The transformer-based language model usually uses an attention mask matrix to deal with the input problem of non-fixed-length sequences, which is formulated as follows:
\begin{equation}
    Q^{i+1}, K^{i+1}, V^{i+1}=h^i W_q, h^i W_k, h^i W_v,
\end{equation}
\begin{equation}
\label{attention}
    h^{i+1}=\operatorname{softmax}\left(\frac{Q^{i+1} K^{{i+1}^{\top}}+M}{\sqrt{d_k}}\right)V^{i+1}.
\end{equation}
Here, $W^q, W^k, W^v$ are trainable parameters of the model, $h^i$ is the hidden state output of the $i$-th layer, $d_k$ is the scaling factor, and $M$ is the attention mask matrix. By designing the attention mask matrix $M$, we can control the attention between input tokens. Therefore, we use the knowledge tree to design the mask matrix $M$, specifically:
\begin{equation}
M_{i j}= \begin{cases}0 & x_i, x_j \in N_o \\ 0 & x_i \in N_1, x_j \in N_2 \\ 0 & x_i \in N_p, x_j \in N_c \text{ or } x_i \in N_c, x_j \in N_p \\ -\infty & \text { otherwise }\end{cases}.
\end{equation}
Here, $x_i$ and $x_j$ are any two tokens in the prompt sequence, $0$ means that the $i$-th token can see the $j$-th token, $-\infty$ means that the $i$-th token cannot see the $j$-th token, $N_m$ represents a node in the knowledge tree, $N_1$ and $N_2$ are nodes with the same parent node, and $N_p$ and $N_c$ are a pair of parent and child nodes with a parent-child relationship in the knowledge tree.

The mask matrix $M$ indicates that, for any node, it can see itself, parent nodes, child nodes, and sibling nodes. For example, the triple prompt $AA_1$ in Figure \ref{fig:visible} can see the entity $A$, triple prompts $AA_2$, $A_1A_{11}$, and $A_1A_{12}$. It can be seen that any node and its visible nodes have the same entity, and invisible nodes contain different entities. Therefore, the knowledge tree mask matrix can keep the original knowledge structure of the knowledge prompt, solve the problem of irrelevant and noisy knowledge, and improve the model's performance.

\subsection{Training and Recommendation}

We employ the T5 model architecture \cite{raffel2020exploring},which is an encoder-decoder-based pre-trained language model using mask prediction as the pre-training task. We construct personalized prompts with masks, transforming the recommendation task into a mask prediction task similar to the pre-training task of PLMs. The loss function is given by:
\begin{equation}
    \mathcal{L}_\theta=-\sum_{j=1}^{|\mathbf{y}|} \log P_\theta\left(\mathbf{y}_j \mid \mathbf{y}_{<j}, \mathbf{X_{prompt}}\right).
\end{equation}
Here, $\theta$ represents the model parameters, and $y$ represents the predicted output.

During the recommendation stage, we use the same method as P5 \cite{geng2022recommendation}, which applies beam search to generate a list of potential next items.

\section{Experiments}

We conduct extensive experiments on three publicly available datasets to answer the following research questions:

\begin{itemize}
\item \textbf{RQ1:} Does KP4SR outperform the current state-of-the-art SR methods?
\item \textbf{RQ2:} Does the conversion of the recommendation task to an NLP task follow the general patterns of the NLP field?
\item \textbf{RQ3:} What is the impact of each component and hyperparameters in KP4SR?
\item \textbf{RQ4:} Does KP4SR have generalization capability for unknown templates?
\end{itemize}

\subsection{Experiments Settings}
\subsubsection{Datasets}

We conduct experiments on three public datasets: Amazon books \cite{he2016ups}, LFM-1b \cite{harper2015movielens}, and Movielens-10M \cite{schedl2016lfm}. These datasets record the interaction information between users and books, music, and movies.

The KG we used is from KB4Rec \cite{Zhao-DI-2019}, which links the above three widely used datasets with the widespread knowledge base Freebase\cite{bollacker2008freebase} to provide side information for recommendation systems. We apply the same preprocessing steps as P5 \cite{geng2022recommendation} and filter out all users and items that appeared less than five times. For the LFM-1b dataset, we filter out tracks that were played less than 10 times. In addition, we search Freebase and convert all entity IDs to text. However, for items, we still use item IDs as inputs and outputs according to the approach in \cite{geng2022recommendation}. The statistics of the preprocessed datasets are shown in Table \ref{tab:dataset}.

\begin{table}[t]
    \caption{Statistics of the datasets.}
    \label{tab:dataset}
    \centering
    \resizebox{0.8\columnwidth}{!}{
    \begin{tabular}{c|c|c|c}
    \hline Stats. & books & music & movies \\
    \hline 
    \# Users & 70,463 & 70,774 & 69,878 \\
    \#Items & 24,921 & 335,619 & 10,130 \\
    \# Interactions & 845,321 & 8,768,434 & 9,991,477 \\
    Density & $4.8 \times 10^{-4}$ & $3.7 \times 10^{-4}$ & $1.4 \times 10^{-2}$ \\
    \hline & \multicolumn{3}{|c}{ Knowledge Graph } \\
    \hline \# Relations & 16 & 6 & 44 \\
    \# Entities & 103,695 & 1,434,216 & 123,306 \\
    \# Triples & 402,543 & 2,724,800 & 954,004 \\
    \hline
    \end{tabular}
    }
\end{table}

\begin{table*}[t]
\centering
\caption{Overall performance. Bold scores represent the highest results of all methods. Underlined scores stand for the second highest results of all methods.}
\label{tab:results}
\resizebox{\textwidth}{!}{
\begin{tabular}{ccccc|cccc|cccc} 
\hline
\multirow{2}{*}{Methods} & \multicolumn{4}{c|}{books}                                                           & \multicolumn{4}{c|}{music}                                                               & \multicolumn{4}{c}{movies}                                                                   \\ 
\cline{2-13}
                         & NDCG@5                & NDCG@10              & HR@5               & HR@10             & NDCG@5                & NDCG@10              & HR@5               & HR@10              & NDCG@5                & NDCG@10              & HR@5               & HR@10                 \\ 
\hline
Caser & 0.0220 & 0.0294 & 0.0356 & 0.0587 & 0.0165 & 0.0232 & 0.0271 & 0.0477 & 0.0309 & 0.0462 & 0.052 & 0.0999 \\
GRU4Rec & 0.0235 & 0.0317 & 0.0380 & 0.0635 & 0.0222 & 0.0317 & 0.0374 & 0.0668 & 0.0378 & 0.0554 & 0.0643 & 0.1191 \\
BERT4Rec & 0.0204 & 0.0282 & 0.0323 & 0.0567 & 0.0242 & 0.0356 & 0.0426 & 0.0781 & 0.0328 & 0.0488 & 0.056 & 0.1062 \\
SASRec & 0.0254 & 0.0362 & 0.0466 & 0.0803 & 0.0327 & 0.047 & 0.0634 & 0.1078 & 0.0312 & 0.0459 & 0.0538 & 0.0994 \\
GRU4RecF & 0.0240 & 0.0321 & 0.0381 & 0.0633 & 0.0266 & 0.0377 & 0.0441 & 0.0788 & 0.0372 & 0.0551 & 0.0644 & 0.1204 \\
GRU4RecKG & 0.0233 & 0.0314 & 0.0373 & 0.0625 & 0.0222 & 0.0313 & 0.0380 & 0.0664 & 0.0374 & 0.0562 & 0.0650 & 0.1237 \\
KSR & 0.0240 & 0.0317 & 0.0383 & 0.0623 & 0.0330 & 0.0411 & 0.0504 & 0.0757 & 0.0394 & 0.0574 & 0.0679 & 0.1242 \\
FDSA & 0.0221 & 0.0309 & 0.0355 & 0.0631 & 0.0185 & 0.0261 & 0.0304 & 0.0539 & 0.0354 & 0.0523 & 0.0604 & 0.1132 \\
SASRecF & 0.0238 & 0.0319 & 0.0379 & 0.0631 & 0.0310 & 0.0418 & 0.0503 & 0.0839 & 0.0294 & 0.0441 & 0.0503 & 0.0964 \\
$\operatorname{S^3Rec}$ & 0.0249 & 0.0356 & 0.0452 & 0.0783 & 0.0301 & 0.0443 & 0.0524 & 0.0968 & 0.0306 & 0.0461 & 0.0536 & 0.1019 \\
DIF-SR & 0.0298 & 0.0416 & 0.0584 & \underline{0.0948} & 0.0573 & 0.0678 & \pmb{0.1110} & \pmb{0.1433} & 0.0492 & 0.0689 & 0.0875 & \pmb{0.1489} \\
\hline
P5 & \underline{0.0433} & \underline{0.0501} & \underline{0.0604} & 0.0813  & \underline{0.0815} & \underline{0.0879} & 0.0994 & 0.1193 & \underline{0.0618}	& \underline{0.0738}  & \underline{0.0888}	& 0.1261  \\ 
KP4SR & \pmb{0.0609} & \pmb{0.0691} & \pmb{0.0824} & \pmb{0.1077} & \pmb{0.0906} & \pmb{0.0975} & \underline{0.1108} & \underline{0.1319} & \pmb{0.0755}	& \pmb{0.0891} & \pmb{0.1058} & \underline{0.1481} \\ 
\hline
\end{tabular}
}
\end{table*}

\subsubsection{Evaluation}

Following the prior work \cite{geng2022recommendation}, we evaluate our KP4SR using Hit Rate@k (HR@k) and Normalized Discounted Cumulative Gain@k (NDCG@k), and report HR@5, 10 and NDCG@5, 10. Higher values indicate better performance for all metrics. We use the leave-one-out strategy to evaluate the performance of each method, which has been widely used in many related works \cite{geng2022recommendation}. Specifically, for each user-item interaction sequence, the last two items are kept as validation and test data, and the rest of the items are used to train the SR model. To make a fair comparison, we evaluate the model performance in a fully ranking manner. The ranking results are obtained on the entire item set rather than sampled results.

\subsubsection{Baselines}

We choose to use the following state-of-the-art SR methods as baselines in our experiments: 

\begin{itemize}
    \item \textbf{Caser} \cite{tang2018personalized}: A CNN-based model that uses horizontal and vertical convolution filters to learn multiple patterns and user preferences.
    \item \textbf{GRU4Rec*}: GRU4Rec \cite{hidasi2015session} is a session-based RS that uses RNNs to capture sequential patterns. GRU4RecF \cite{hidasi2016parallel} incorporates item attribute information. GRU4RecKG is an extension of GRU4Rec that connects items and their corresponding KG embeddings as inputs.
    \item \textbf{BERT4Rec} \cite{sun2019bert4rec}: A bidirectional self-attention network that models user behavior sequences using cloze tasks.
    \item \textbf{SASRec*}: SASRec \cite{kang2018self} is an attention-based model that uses self-attention networks for SR. SASRecF is an extension of SASRec that connects items and their features as inputs.
    \item \textbf{KSR} \cite{huang2018improving}: An RNN and memory-based model that captures attribute-level user preferences using KG.
    \item \textbf{FDSA} \cite{zhang2019feature}: A self-attention-based model that integrates heterogeneous features into a feature sequence for SR.
    \item \boldmath$\operatorname{S^3Rec}$\unboldmath \, \cite{ zhou2020s3}: A self-supervised learning-based model that has four carefully designed optimization objectives for learning correlations in raw data.
    \item \textbf{DIF-SR} \cite{xie2022decoupled}: An attention-based model that performs sequence recommendation by decoupling side information fusion.
    \item \textbf{P5} \cite{geng2022recommendation}: A PLM-based recommendation system that unifies multiple recommendation tasks into a single framework through personalized prompt sets. 
\end{itemize}

\subsubsection{Implementation Details}

Our KP4RS utilizes a pre-trained T5 model as its backbone. For KP4RS, the encoder and decoder have six layers, a model dimension of 512, and 8 heads of attention. During the training phase, we use mixed precision to accelerate the training speed. We use 4 Ascend 910 NPUs with a batch size of 64. We use the AdamW optimizer \cite{loshchilov2017decoupled} with a peak learning rate of 1e-3 and set the maximum length of the input tokens to 512. Due to the length limitation of the input (see Appendix for details), we only study the knowledge prompts within three hops.

For all baselines, if negative samples are needed to calculate the next item prediction loss during the training phase, we follow the usual practice of randomly selecting a negative sample for each interaction. For a fair comparison, the maximum interaction sequence length for all experiments is set to 5 unless otherwise specified.

\subsection{Overall Performance Comparison (RQ1)}


The results of different methods on three datasets are shown in Table \ref{tab:results}. According to the experimental results, we can see that, compared to the five SR methods (Caser, GRU4Rec, BERT4Rec, SASRec, and P5) that do not integrate side information, P5 outperforms other methods by a large margin. We believe there are two main reasons for this: firstly, P5 is a PLM-based method with better initialization parameters and a larger model that can accommodate more information. Secondly, P5 converts the recommendation task into an NLP task, which can leverage the general knowledge in PLMs. This indicates that PLMs have great potential for application in solving recommendation tasks.
SASRec ranks second to P5 and performs well on books and music datasets, but it underperforms on the movie dataset. This suggests that different datasets have varying sequence patterns, impacting the performance of SR methods.

Some SR methods that integrate side information do not achieve better performance. For example, GRU4RecF and GRURecKG fuse item attribute information and KG information at an early stage, and they do not achieve better results. This is because early feature fusion may result in the loss of useful information. Similarly, SASRecF directly connects items and item features as input, and they also suffer from the information loss issue. In contrast, DIF-SR achieves much better results than other baselines. This is because DIF-SR decouples the fusion of side information, moves the side information from input to attention layers, and further decouples the attention calculation of various side information and item representations.

Finally, it is clear that our KP4SR achieves the best results in most evaluation metrics on all three datasets, especially outperforming other methods by a large margin on the NDCG metric. KP4SR and P5 use the same PLM architecture, but our method performs much better than P5 on all datasets. Specifically, for NDCG@5 and HR@5, our method achieved a 40.65\% and 36.42\% improvement on the books dataset, an 11.17\% and 11.47\% improvement on the music dataset, and a 22.17\% and 19.14\% improvement on the movies dataset. This is because KP4SR not only converts the recommendation task into a pre-training task but also efficiently utilizes domain knowledge.

\begin{table}[t]
\centering
\caption{Pre-training fine-tuning vs Prompt-tuning}
\label{tab:finetune_prompt}
\resizebox{\columnwidth}{!}{
\begin{tabular}{ccccccc} 
\hline
                       & method      & NDCG@5  & NDCG@10 & HR@5   & HR@10  & epoch     \\ 
\hline
\multirow{3}{*}{books}  & P5        & 0.0433  & 0.0501  & 0.0604 & 0.0813 & 300       \\
                       & MPP         & 0.0505  & 0.0570  & 0.0658 & 0.0861 & 90        \\ 
\cline{2-7}
                       & Improvement & 16.63\% & 13.77\% & 8.94\% & 5.90\% & -70\%     \\ 
\hline
\multirow{3}{*}{music} & P5        & 0.0815  & 0.0879  & 0.0994 & 0.1193 & 300       \\
                       & MPP         & 0.0839  & 0.0904  & 0.1018 & 0.1220 & 140       \\ 
\cline{2-7}
                       & Improvement & 2.94\%  & 2.84\%  & 2.41\% & 2.26\% & -53.33\%  \\ 
\hline
\multirow{3}{*}{movies} & P5        & 0.0618	& 0.0738  & 0.0888	& 0.1261      & 260       \\
                       & MPP         & 0.0645  & 0.0765  & 0.0906 & 0.1282 & 150       \\ 
\cline{2-7}
                       & Improvement & 4.37\%   & 3.66\%   & 2.03\%    & 1.67\%      & -42.31\%  \\
\hline
\end{tabular}}
\end{table}

\subsection{Pre-training fine-tuning vs Prompt-tuning (RQ2)}

P5 uses a collection of personalized prompt templates to transform recommendation tasks into question-answering tasks. This differs from pre-training tasks and essentially belongs to the pre-training fine-tuning paradigm. We design the MPP, which can transform a recommendation task into a pre-training task, specifically a cloze task, as shown in Table \ref{tab:finetune_prompt}. It can be observed that MPP achieves better performance on all three datasets and converges faster, reducing the number of training epochs by 70\%, 53.33\%, and 42.31\%, respectively. Therefore, compared to first pre-training and then fine-tuning, transforming recommendation tasks into NLP tasks and then using prompt-tuning for recommendation is more competitive.




\subsection{The impact of KP and PD (RQ3)}
\label{KP_KTM}
Next, we conduct experiments to investigate the impact of KP and PD on three datasets. The results are reported in Table \ref{tab:KP_KTM}.
From the table, we can observe:

Firstly, without PD, the model performance gradually improves with the increase of knowledge prompt hops on books and music datasets. On the movies dataset, the best performance is achieved with 1-hop knowledge prompts, and too much knowledge can negatively affect performance. This suggests that PLMs can leverage knowledge prompts to help capture user preferences and improve recommendation performance, but the impact may vary across different datasets.

Secondly, when the KP hops are the same, PD generally improves performance, indicating that the proposed PD can effectively mitigate knowledge noise. However, PD is not always effective as it may ignore potential relationships between triple prompts.

\begin{table}[t]
\centering
\caption{Results when using different settings of KP and PD. $n$ represents the number of KP hops.}
\label{tab:KP_KTM}
\resizebox{\columnwidth}{!}{
\begin{tabular}{cc|cc|cc|cc} 
\hline
\multicolumn{2}{c|}{ setting} & \multicolumn{2}{c|}{books} & \multicolumn{2}{c|}{music} & \multicolumn{2}{c}{movies}  \\
KP & PD              & NDCG@5 & HR@5             & NDCG@5 & HR@5              & NDCG@5 & HR@5              \\ 
\hline
0  & $\boldsymbol{-}$ & 0.0505      & 0.0658       & 0.0839      & 0.1018                 & 0.0645      & 0.0906                 \\
1  & \usym{2717}    & 0.0503      & 0.0697       & 0.0881      & 0.1077                 & 0.0671      & 0.0944                 \\
2  & \usym{2717}    & 0.0522      & 0.0722       & 0.0885      & 0.1080                 & 0.0650      & 0.0923                 \\
3  & \usym{2717}    & 0.0543      & 0.0753       & 0.0891      & 0.1093                 & 0.0623      & 0.0892                 \\
1  & \usym{2713}      & 0.0542     & 0.0734       & 0.0894      & 0.1091                 & \pmb{0.0729}      & \pmb{0.1012}                 \\
2  & \usym{2713}      & \pmb{0.0575}      & \pmb{0.0787}       & 0.0886      & 0.1086                 & 0.0677      & 0.0959                 \\
3  & \usym{2713}      & 0.0523      & 0.0737       & \pmb{0.0899}      & \pmb{0.1102}                 & 0.0702      & 0.0980                 \\
\hline
\end{tabular}}
\end{table}

\subsection{Impact of the degree of knowledge tree (RQ3)}

We investigate the influence of the degree of the knowledge tree (i.e., the number of tail entities connected to the head entity) on recommendation performance. The results are shown in Table \ref{tab:degree}. On the books dataset, the performance improves as the degree increases, while on the music and movies datasets, the performance decreases after a certain degree. This phenomenon is consistent with the analysis in Section \ref{KP_KTM}, which suggests that a certain amount of knowledge can improve performance, but excessive noise knowledge may lower performance.

\begin{table}[t]
\centering
\caption{Performance with different degrees of knowledge tree. *For the books dataset, the degrees are 2, 4, 6, and 8.}
\label{tab:degree}
\resizebox{\columnwidth}{!}{
\begin{tabular}{c|cc|cc|cc} 
\hline
\multirow{2}{*}{degrees} & \multicolumn{2}{c|}{books} & \multicolumn{2}{c|}{music} & \multicolumn{2}{c}{movies}  \\
                        & NDCG@5 & HR@5             & NDCG@5 & HR@5              & NDCG@5 & HR@5              \\ 
\hline
2*                      & 0.0575      & 0.0787                & 0.0890      & 0.1097                 & 0.0729      & 0.1012                 \\
3*                      & 0.0588      & 0.0807                & 0.0889      & 0.1097                 & 0.0738      & 0.1032                 \\
4*                      & 0.0608      & 0.0813                & \pmb{0.0906}      & \pmb{0.1108}                 & \pmb{0.0755}      & \pmb{0.1058}                 \\
5*                      & \pmb{0.0609}      & \pmb{0.0824}                & 0.0899      & 0.1102                 & 0.0712      & 0.1005                 \\
\hline
\end{tabular}}
\end{table}

\subsection{Maximum sequence length}
\label{item_length_ex}

\begin{figure}[t]
  \includegraphics[width=\columnwidth]{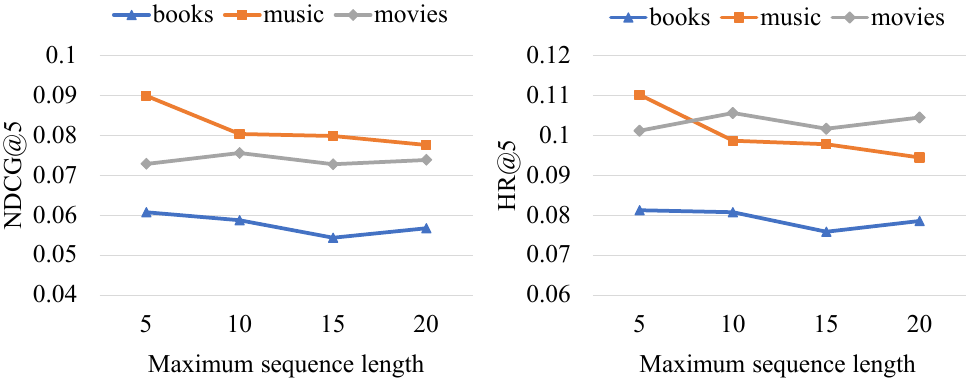}
  \caption{Performance with different maximum sequence length.}
  \label{fig:item_length}
\end{figure}

Due to the input length limitation, the maximum length of the user-item interaction sequence in the previous studies was set to 5. In this section, we investigate the effect of maximum interaction sequence length on recommendation performance, as shown in Figure \ref{fig:item_length}. On the music dataset, the performance decreases as the maximum sequence length increases. On the books and movies datasets, the maximum sequence length has little impact on performance. This indicates that users' behavior is more dependent on their recent interactions with items, and a larger maximum sequence length may not necessarily lead to better performance, as it may introduce additional noise.

\subsection{Generalization of KP4SR (RQ4)}

\begin{figure}[t]
  \includegraphics[width=\columnwidth]{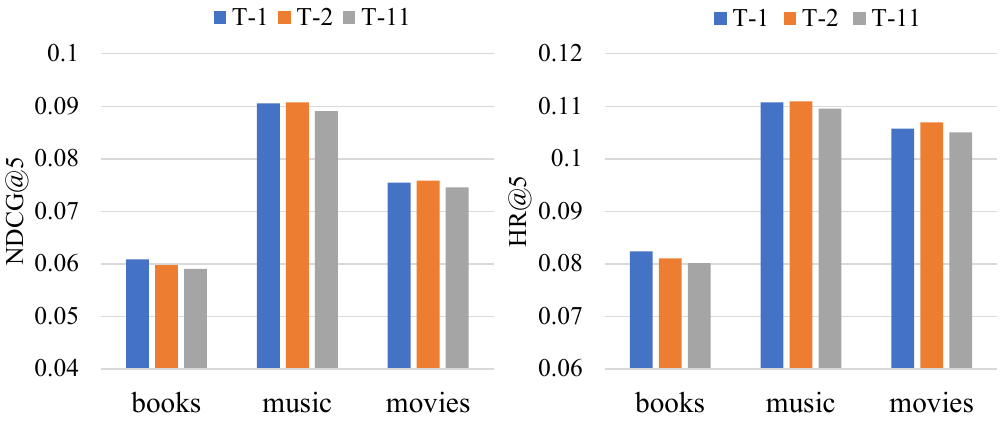}
  \caption{Performance with different prompt templates.}
  \label{fig:generalization}
\end{figure}

We manually design 11 MPP templates for each dataset, with the first ten used for training, the first one used as the default testing template, and the eleventh used to test the model's generalization to unknown templates. We evaluate the performance of three templates, and the results are shown in Figure \ref{fig:generalization}. Comparing templates 1 and 2, we can see that there are performance differences across different templates, indicating that better-designed templates can lead to better performance. At the same time, the model's performance only slightly decreases on unknown templates, indicating that KP4SR has a good generalization ability to unknown templates.

\section{Conclusion and future work}

In this paper, we propose KP4SR, which is the first work that transforms KG into knowledge prompts to improve SR. Firstly, we design a set of MPP templates to transform the SR task into an NLP task, significantly improving the recommendation performance and convergence speed. Secondly, we construct a set of relation templates to transform the KG into KP. It not only addresses the problems of semantic differences but also enables the easy utilization of high-order information in the KG. Next, we propose PD to alleviate noise issues. The PD restores the data structure in the form of a mask matrix, which eliminates the impact of irrelevant triples. Finally, extensive experiments demonstrate the effectiveness of our method.
In future work, we will continue to explore how to use large language models to achieve better recommendation performance.

\begin{acks}

This work was supported by Key-Area Research and Development Program of Guangdong Provinc (2021B0101400002), NSFC (62276277), Guangdong Basic and Applied Basic Research Foundation (2022B1515120059), Guangdong Provincial Key Laboratory of Intellectual Property and Big Data (2018B030322016), The Major Key Project of PCL(PCL2021A13).
\end{acks}

\bibliographystyle{ACM-Reference-Format}
\balance
\bibliography{sample-new}

\newpage 
\appendix

\section{Length of input tokens}
\label{Length of input tokens}
\begin{figure}[b]
	\centering
	\subfigure[books]{
		\label{book_length}
		\includegraphics[width=0.46\columnwidth]{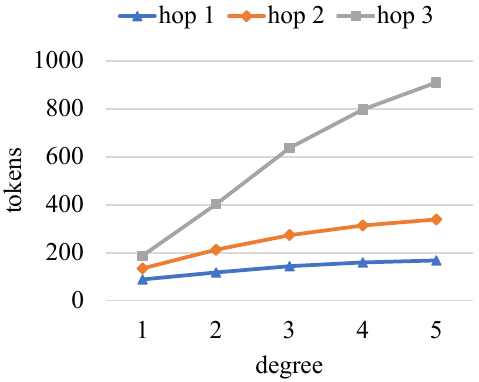}} 
	\subfigure[music]{
		\label{music_length}
		\includegraphics[width=0.46\columnwidth]{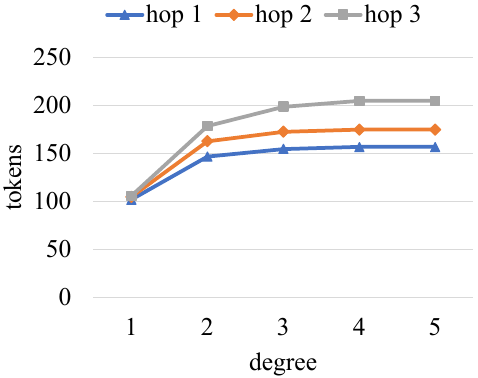}}

 	\subfigure[movies]{
		\label{movie_length}
		\includegraphics[width=0.46\columnwidth]{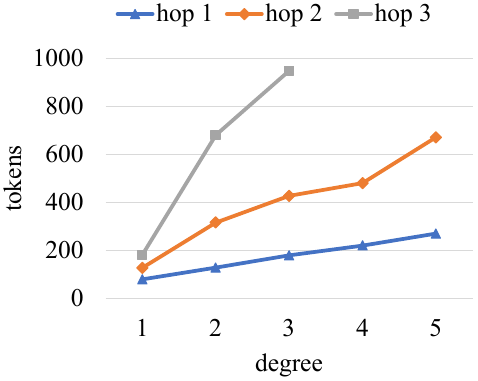}} 
	\subfigure[items]{
		\label{items_length}
		\includegraphics[width=0.46\columnwidth]{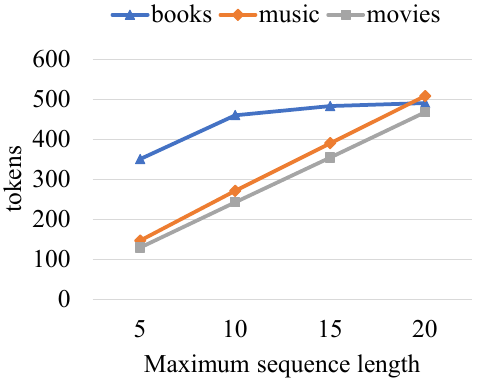}}
	\caption{Length of input tokens}
	\label{appendix_length}
\end{figure}

We statistic the input tokens length of three datasets under different conditions, as shown in Figure \ref{appendix_length}. Each data point represents the median length of all samples rather than the mean. For KP4SR, the default maximum input length is 512. Therefore, in Section 5.3, we set the degree to 2 for the books and movies datasets, and set the degree to 5 for the music dataset. In Section 5.6, we set the hop to 2 and the degree to 6 for the books dataset, and set the hop to 1 and the degree to 2 for the music and movies datasets. Figure \ref{items_length} shows the length of input samples under different maximum sequence lengths.

\end{document}